\documentclass[letterpaper, 10 pt, conference]{ieeeconf}  

\IEEEoverridecommandlockouts                              
\usepackage{graphicx} 
\usepackage{amsfonts}
\usepackage{url}
\usepackage{amsmath}
\usepackage{amssymb}
\usepackage{xcolor}
\usepackage{algorithm}
\usepackage{microtype}
\usepackage{graphicx}
\usepackage{subcaption}
\usepackage{booktabs}
\usepackage{algorithmic}
\usepackage{booktabs}
\usepackage{mathtools}
\mathtoolsset{showonlyrefs}
\usepackage[dvipsnames]{xcolor}

\usepackage{multirow}
\newcommand{\Up}{U_{\mathrm{p}}}
\newcommand{\Yf}{Y_{\mathrm{f}}}
\newcommand{\Yfh}{\hat{Y}_{\mathrm{f}}}
\newtheorem{theorem}{Theorem}
\newtheorem{proposition}[theorem]{Proposition}

\title{\LARGE \bf
Dynamics Aware Past-Future Information Bottleneck
}

\author{Wooyoung Chung$^{1}$ and Stas Tiomkin$^{1}$
\thanks{$^{1}$The authors are with Texas Tech University, Lubbock, TX 79409, USA. 
{\tt\small \{wochung, stas.tiomkin\}@ttu.edu}}
}

\usepackage{hyperref}
\hypersetup{
    colorlinks=true,
    linkcolor=blue,
    citecolor=red
}

\begin{document}

\maketitle
\thispagestyle{empty}
\pagestyle{empty}

\begin{abstract}
The simplification of complex dynamical systems is essential for efficient estimation and control. A principled approach to this task is through the lens of information-theoretic compression. The Past–Future Information Bottleneck (PF-IB) provides a theoretical framework for compressing linear dynamical systems. However, the standard formulation often lacks structural consistency, yielding representations that are not realizable as dynamical systems. By constraining the objective to a specific manifold of block Hankel matrices, which we prove is a Lie group, we bridge the gap between PF-IB and realization theory. Our approach ensures that the resulting compressed dynamics are realizable, thereby enabling its direct application to estimation and control. We demonstrate our approach in numerical simulations.

\end{abstract}

\section{INTRODUCTION}
Managing high-dimensional dynamical systems remains a foundational challenge in control \cite{bellman1957dynamic}: as systems grow in scale, training, estimation, and control become increasingly difficult \cite{siljak1991decentralized,tsitsiklis1985complexity}, making system simplification essential. One principled approach to simplification is through compression.

The Information Bottleneck (IB) \cite{Tishbyetal1999,slonim2002information,tishby2011information} formalizes compression as minimizing representation complexity while maximizing the information retained about a target output. PF-IB \cite{Creutzig2009} applies this principle to linear dynamics, treating the system's past controls as the input and its future observations as the output to be predicted. However, standard PF-IB optimizes over an unstructured space, and the resulting compressed representations often fail to correspond to a valid dynamical system. This creates a realization gap, necessitating heuristic approximation of a dynamical system from the bottleneck's representations \cite{Creutzig2009,AHMAD2024105150}.


To bridge this gap, the compressed representation must respect the underlying structure of a proper dynamical system. Classically, the past-future relation is captured through the Hankel matrix~\cite{Kalman1960}, from which the Ho--Kalman construction~\cite{HoKalman1966} recovers a realization. A rich literature builds on this foundation: balanced truncation and Hankel-norm approximation~\cite{moore1981principal,glover1984all} produce reduced realizations with norm-based error guarantees, the Loewner framework~\cite{mayo2007framework} constructs realizations directly from data, and data informativity~\cite{vanwaarde2020data} characterizes when finite data suffices for identification and control. These methods are realizable by construction but optimize norm-based criteria at fixed model order, whereas PF-IB's continuous compression--prediction tradeoff has historically come at the cost of realizability. We propose that the compressed operator must instead preserve the Hankel structure, ensuring realizability by construction~\cite{HoKalman1966}.

In this paper, we formalizing the dynamical system structure as a Lie group, mapping the information bottleneck objective directly onto the natural geometric manifold of the system \cite{jurdjevic1997geometric,celledoni2025model}. This guarantees that the bottleneck output is inherently a valid dynamical system, eliminating the need for post-hoc approximation, and enables more accurate state estimation and more stable control performance.

Our main contributions are as follows:
\begin{itemize}
\item \textbf{Extension to Block Hankel Manifolds:} We extend the standard Hankel Lie algebra to the block Hankel manifold where the elements are matrices of the General Linear Group. This extension enables the compression of Multi-Input Multi-Output (MIMO) systems and allowing for more complex, curved optimization.
\item \textbf{Structured PF-IB Optimization:} We develop a novel optimization framework for the PF-IB objective that constrains the solution to the block Hankel manifold.
\item \textbf{Empirical Validation in Estimation and Control:} We demonstrate that our method addresses the realization gap, outperforming standard PF-IB in both state estimation and LQR control tasks.
\end{itemize}

The paper is organized as follows. Sec. \ref{sec:pfib} reviews the general concept of linear systems and Past-Future Information Bottleneck. Sec. \ref{sec:theoretical_motivation} provides the theoretical motivation by connecting the recently proven standard Hankel manifold \cite{AHMAD2024105150} to dynamical systems. Sec. \ref{sec:blockhankel} then extends this theory to the block Hankel manifold, providing the necessary components to bridge the gap for high-dimensional systems. Sec. \ref{sec:riemmanianpfib} utilizes the block Hankel manifold to formulate our realizable PF-IB objective. Sec. \ref{sec:results} demonstrates the closure of the realization gap and evaluates performance on estimation and control tasks. Sec. \ref{sec:conclusion} analyzes the broader impact of this approach, discusses limitations, and suggests future research directions. Proofs of the results presented in Sec. \ref{sec:blockhankel} can be found in the Appendix (Sec. \ref{sec:Appendix}).

\section{PAST-FUTURE INFORMATION BOTTLENECK}\label{sec:pfib}

\subsection{Linear Dynamical Systems and the Hankel Representation}
We begin by reviewing the representation of linear dynamical systems through their past–future structure, which forms the foundation of PF-IB.

Consider a discrete-time, stable, linear time-invariant (LTI) system
\begin{align}
x_{t+1} &= A x_t + B u_t, \\
y_t &= C x_t + D u_t,
\end{align}
where $u_t$, $x_t$. and $y_t$ are $p \times 1$, $m \times 1$, and $q \times 1$ vectors and $A$, $B$, $C$, and $D$ are $m \times m$, $m \times p$, $q \times m$, and $q \times p$ matrices.

Given the past input sequence $\Up$ and the future output sequence
$\Yf$ over a finite horizon of $k$ steps,
\begin{equation}
\Up = \begin{bmatrix} u_{t} \\ u_{t-1} \\ \vdots \\ u_{t-k+1} \end{bmatrix},
\quad
\Yf = \begin{bmatrix} y_{t+1}\\ y_{t+2} \\ \vdots \\ y_{t+k}
\end{bmatrix},
\end{equation}
where, following the PF-IB setting~\cite{Creutzig2009}, future inputs
are set to zero ($u_\tau = 0$ for $\tau > t$), so $\Yf$ is the natural
response induced by the past inputs.

With controllability matrix
$\mathcal{C}_k$ and the observability matrix $\mathcal{O}_k$ defined as

\begin{equation}
\mathcal{C}_k =
\begin{bmatrix}
B & AB & \cdots & A^{k-1}B
\end{bmatrix},
\qquad
\mathcal{O}_k =
\begin{bmatrix}
C \\
CA \\
\vdots \\
CA^{k-1}
\end{bmatrix},
\end{equation}
the mapping from past inputs $\Up$ to future outputs $\Yf$ is

\begin{equation}\Yf = \mathcal{O}_k \mathcal{C}_k \Up \equiv H \Up.\end{equation}

The operator $H$ forms a \textbf{block Hankel matrix}: under its
partition into a $k \times k$ grid of $q \times p$ blocks, the
$(i,j)$-th block is the Markov parameter (impulse response
parameter)
\begin{equation}\label{eq:markov}
    G_i = C A^{i-1} B, \qquad i = 1,2,\dots
\end{equation}
arranged, under the partition into a $k \times k$ grid of
$q \times p$ blocks, so that the $(i,j)$-th block is $G_{i+j-1}$,
constant along each anti-diagonal:
\begin{equation} \label{eq:hankelmatrix}
H =
\begin{bmatrix}
G_1 & G_2 & \cdots & G_k \\
G_2 & G_3 & \cdots & G_{k+1} \\
\vdots & \vdots & \ddots & \vdots \\
G_k & G_{k+1} & \cdots & G_{2k-1}
\end{bmatrix}.
\end{equation}

This representation is central to PF-IB, as it provides a natural domain in which to reason about predictive structure and realizability.

\subsection{Past-Future Information Bottleneck}\label{subsection:pfib}



PF-IB seeks a representation of the past of a stochastic process
that is maximally informative about its future while remaining as
compact as possible. The information bottleneck objective can be
written as
\begin{equation}
\min \; \mathcal{L} \;=\; I(\Up;\Yfh) \;-\; \beta\, I(\Yfh;\Yf),
\label{eq:ib}
\end{equation}
where $\beta>0$ controls the compression--prediction tradeoff and $I(\cdot\,;\cdot)$ denotes mutual information. Under Gaussian assumptions, PF-IB yields a compressed representation
\begin{equation}
\Yfh = H(\beta) \Up + \varepsilon,
\end{equation} where $H(\beta)$ is the compressed linear operator obtained from the original Hankel matrix $H$ at trade-off level $\beta$, and $\varepsilon \sim \mathcal{N}(0,\Sigma_\varepsilon)$ is additive Gaussian noise. The objective Eq.~\eqref{eq:ib} then becomes
\begin{multline}
    \mathcal{L}= (1-\beta)\,\log\!\big \lvert H(\beta)\Sigma_{\Up}H(\beta)^\top + \Sigma_\varepsilon\big \rvert - \log \!\big \lvert \Sigma_\varepsilon \!\big \rvert \\+\beta \log \!\big \lvert H(\beta)\Sigma_{\Up|\Yf}H(\beta)^\top + \Sigma_\varepsilon\big \rvert \label{eq:pfib_logdet}
\end{multline}
where $\Sigma_{\Up}$ is the covariance of the past inputs and
$\Sigma_{\Up|\Yf}$ is the conditional covariance. As shown in \cite{Chechiketal2005}, without loss of generality the noise covariance can be normalized to $\Sigma_\varepsilon = I$. Differentiating Eq. \eqref{eq:pfib_logdet} with respect to $H(\beta)$ yields


\begin{equation}  \label{eq:pfib_gradient}
\begin{aligned}[b]
\frac{d\mathcal{L}}{dH}
  = {}& (1-\beta)\bigl[H\Sigma_{U_p}H^{\top} + I\bigr]^{-1} 2H\Sigma_{U_p} \\
  {}+{}& \beta\bigl[H\Sigma_{U_p|Y_f}H^{\top} + I\bigr]^{-1} 2H\Sigma_{U_p|Y_f}.
\end{aligned}
\end{equation}

While the PF-IB objective admits a closed-form solution via a
weighted singular value decomposition~\cite{Creutzig2009,
Chechiketal2005}, this derivation is fundamentally unconstrained,
which opens a structural gap.

\textbf{Realizability Gap:} As explained in Sec. \ref{sec:pfib}, the past--future operator of any LTI system is block Hankel: its $(i,j)$ block is the Markov parameter $G_{i+j-1}$, so entries along each anti-diagonal coincide. The closed-form optimum, however, treats every entry of the operator as a free variable and thus generically violates these anti-diagonal constraints. The resulting $H(\beta)$, while information-theoretically optimal, is not the past--future map of \emph{any} state-space system, of any order. Extracting $(A,B,C)$ via Ho--Kalman~\cite{Hokalman} therefore amounts to projecting $H(\beta)$ onto the set of realizable operators, and this projection discards predictive information. Standard PF-IB thus faces a conflict between information-theoretic optimality and physical consistency: the optimization drifts off the manifold of valid systems and must be heuristically forced back. Our approach instead constrains every iterate to the block Hankel manifold, so the optimum is realizable by construction and no projection is needed.

\section{THEORETICAL MOTIVATION} \label{sec:theoretical_motivation}
To constrain the PF-IB optimization to realizable systems, we seek
a geometric structure on the set of Hankel matrices: if this set
forms a Lie group, optimization can be carried out directly on the
manifold, with updates guaranteed to remain realizable. This
section reviews the recently established group structure for the
scalar case and shows why it must be extended for our setting.

\subsection{Scalar Hankel Manifolds and SISO Realization} \label{sec:theoretical_motivation_scalar}

Recent mathematical developments have characterized the set
$\mathcal{H}$ of $k \times k$ scalar Hankel matrices,
\begin{equation} \label{eq:scalarhankel}
\mathcal{H} = \left\{
\begin{bmatrix}
a & b & \cdots & f \\
b & c & \cdots & g \\
\vdots & \vdots & \ddots & \vdots \\
f & g & \cdots & h
\end{bmatrix}
\;\middle|\;
a, b, \dots, h \in \mathbb{R}
\right\},
\end{equation} as a Lie group under the scalar Hadamard (element-wise)
product~\cite{AHMAD2024105150}.

\begin{proposition}
The set of Hankel matrices under the Hadamard product forms a Lie group over $\mathbb{R}_{+}$. \cite{AHMAD2024105150}
\end{proposition} 

In this setting, the entries $G_i$ (Eq.~\eqref{eq:markov}) are the
scalar Markov parameters of the system's impulse response, so the
Lie group structure renders the space of realizable SISO systems a
smooth manifold, where moving along the manifold corresponds to a
structured evolution of the system dynamics. This connection is, to
our knowledge, new: the classical geometry of minimal linear
systems of fixed McMillan
degree~\cite{brockett1976some,hazewinkel1980fine} concerns
quotients of realizations under basis change, whereas the Hadamard
structure of~\cite{AHMAD2024105150} is a group operation on the
Hankel data itself.

\begin{proposition}
The vector space $\mathfrak{h}$ of Hankel matrices over $\mathbb{R}$ with commutator operation $[A,B] = A \circ B -B \circ A =  0$, where $\circ$ denotes the Hadamard pairwise component product, is a $2k - 1$ dimensional Lie algebra. \cite{AHMAD2024105150}
\end{proposition}

Because scalar multiplication is commutative, $\mathcal{H}$ is an
Abelian Lie group~\cite{hall2015lie}, and the manifold is flat, with zero intrinsic curvature. While this makes
the exponential map $\exp_{\mathcal{H}}: \mathfrak{h} \to
\mathcal{H}$ globally well-behaved, the group structure itself is trivial in
the sense that all Lie brackets vanish.

\subsection{Need for Extension for MIMO Systems} \label{sec:theoretical_motivation_extension}
The scalar Hankel manifold $\mathcal{H}$ applies only to SISO
systems, and the obstruction to extending it lies in the group
operation itself: the scalar theory combines scalar Markov
parameters through the Hadamard product $\circ$, whereas the Markov
parameters of a MIMO system are matrix blocks
(Eq.~\ref{eq:markov}), whose natural entry-wise operation is
block-wise matrix multiplication. Since matrix multiplication does
not commute, the resulting structure is non-Abelian and its
manifold curved. We emphasize that this curvature is a property of
the \emph{space of systems}, not of the systems themselves; every
system considered remains linear. Extending the Lie theory to MIMO
systems is the subject of the next section.

\section{BLOCK HANKEL MANIFOLD}\label{sec:blockhankel}

In this section, we present our core theoretical contribution: the
Lie-algebraic formulation for the block Hankel manifold, extending
the scalar theory of Sec.~\ref{sec:theoretical_motivation} to the
matrix-block setting required for MIMO systems.

\subsection{Topology and Geometry of Block Hankel Matrix}
The space of square MIMO systems with invertible Markov parameters
is characterized by restricting the block Hankel matrices of
Eq.~\ref{eq:hankelmatrix}, with $p = q \equiv n$, to the General Linear Group
$GL(n,\mathbb{R})$:
\begin{equation}
\mathcal{H}_{GL} = \left\{ H \;\middle|\; H_{i,j} = G_{i+j-1},\;
G_m \in GL(n,\mathbb{R}) \right\}.
\end{equation} This restriction provides the manifold with a smooth topology and a
well-defined group inverse at every point; we show that
$\mathcal{H}_{GL}$ forms a Lie group under the block Hadamard
product (Appendix~\ref{sec:Appendix}).

\subsubsection{Group Operator: Block Hadamard Product}
The group operation on $\mathcal{H}_{GL}$ is the block Hadamard
product $\odot$, defined for any $A, B \in \mathcal{H}_{GL}$ by
\begin{equation} \label{eq:blockhadamard}
    (A \odot B)_{i,j} = A_{i,j} B_{i,j}.
\end{equation}
Here $A_{i,j}B_{i,j}$ is ordinary matrix multiplication of the
$n \times n$ blocks, under the same partition as
Eq.~\ref{eq:hankelmatrix}; in contrast to the scalar Hadamard
product $\circ$ of Sec.~\ref{sec:theoretical_motivation}, these
block products do not commute.

\subsubsection{Exponential and Logarithmic Mapping}
The transition between the Lie algebra $\mathfrak{h}_{GL}$ and the
manifold $\mathcal{H}_{GL}$ is governed by block-wise mappings,
owing to the product structure of the group: the map
$H \mapsto (G_1,\dots,G_{2k-1})$ is a Lie group isomorphism onto
the direct product $GL(n,\mathbb{R})^{2k-1}$, under which
$\mathfrak{h}_{GL} \cong \bigoplus_{m=1}^{2k-1}
\mathfrak{gl}(n,\mathbb{R})$ with the block-wise commutator acting
component-wise~\cite{Hall2015,Lee2012}. The Lie-group exponential
of $(\mathcal{H}_{GL},\odot)$ therefore acts block-wise, applying
the standard matrix exponential $\exp(\cdot)$ to each block: for
$h \in \mathfrak{h}_{GL}$,
\begin{equation}\label{eq:exp_map}
\left[ \exp_{\mathcal{H}_{GL}}(h) \right]_{i,j} = \exp(h_{i,j}),
\end{equation} a smooth local diffeomorphism at the identity. Conversely, the
logarithmic map
$\log_{\mathcal{H}_{GL}}: \mathcal{H}_{GL} \to \mathfrak{h}_{GL}$
applies the standard matrix logarithm $\log(\cdot)$ block-wise:
\begin{equation}\label{eq:log_map}
\left[ \log_{\mathcal{H}_{GL}}(A) \right]_{i,j} = \log(A_{i,j}).
\end{equation}

\subsubsection{Projection to the Lie Algebra $\mathfrak{h}_{GL}$}
Constraining the optimization to the manifold requires mapping the
Euclidean gradient $\nabla \mathcal{L}$ into the Lie algebra
$\mathfrak{h}_{GL}$: the unconstrained gradient treats every entry
of the block matrix as an independent variable, so Euclidean
updates immediately exit the Hankel subspace. We therefore define the projection
$\mathcal{P}_{\mathfrak{h}_{GL}}: \mathrm{Mat}(nk,\mathbb{R}) \to
\mathfrak{h}_{GL}$. For the block Hankel structure, this projection
is the block-wise anti-diagonal averaging
\begin{equation}
\left[ \mathcal{P}_{\mathfrak{h}_{GL}}(M) \right]_{i,j} =
\frac{1}{N_{i+j}} \sum_{p+q = i+j} M_{p,q},
\label{eq:projection_operator}
\end{equation}
where $M_{p,q}$ denotes the $(p,q)$-th block of $M$ and
$N_{d} = \min(d-1,\, 2k-d+1)$, $d = 2, \dots, 2k$, is the number
of blocks on the $d$-th anti-diagonal. Applying this operator to
the Euclidean gradient yields the Riemannian gradient
\begin{equation}\label{eq:proj}
\text{grad}\,\mathcal{L}(H(\beta)) =
\mathcal{P}_{\mathfrak{h}_{GL}}(\nabla \mathcal{L}).
\end{equation}
Note that while the anti-diagonal averaging is linear, the
resulting update on the manifold via
$\exp_{\mathcal{H}_{GL}}(\eta \cdot \text{grad}\,\mathcal{L})$ is
highly non-linear \emph{as a map}: the optimization leverages the
linear simplicity of the Lie algebra while respecting the curved
geometry of the MIMO realization space.

\subsection{Connection to MIMO System}
The transition from the scalar manifold $\mathcal{H}$ to the block
manifold $\mathcal{H}_{GL}$ is a fundamental shift in optimization
geometry: the commutative scalar products of $(\mathcal{H}, \circ)$
render it Abelian and flat, whereas the blocks of $\mathcal{H}_{GL}$
multiply non-commutatively, so the Lie bracket
$[A, B] = A \odot B - B \odot A \neq 0$ and the manifold is
intrinsically curved.

The restriction to $GL(n, \mathbb{R})$ that makes this structure a
rigorous Lie group with smooth inverses also imposes the constraint
that every block remain square and full-rank. At first glance, this
appears to conflict with the traditional view of PF-IB, where
compression is achieved through the collapse of singular values.

Strictly speaking, constraining the blocks to $GL(n, \mathbb{R})$
means the optimization runs over full-rank block Hankel operators
only, while for compressive $\beta$ the best block Hankel operator
may be rank-deficient. This exclusion costs nothing: any
rank-deficient operator can be approximated arbitrarily well by
full-rank ones (perturbing each block
$G_m \mapsto G_m + \epsilon I$ preserves the Hankel structure and
is invertible for all but finitely many $\epsilon$), and the IB objective is continuous, so full-rank
operators achieve objective values arbitrarily close to the
rank-deficient optimum. The benefit of remaining full-rank is that
every iterate is exactly realizable.

This realizability is guaranteed by classical partial realization
theory~\cite{Hokalman, Tether1970}: an element of
$\mathcal{H}_{GL}$ is determined by its $2k-1$ anti-diagonal blocks
$(G_1, \dots, G_{2k-1})$, and any finite matrix sequence admits a
finite-dimensional realization $(A, B, C)$ with $CA^{i-1}B = G_i$,
$i = 1, \dots, 2k-1$. Thus every point on the manifold corresponds
to the past--future operator of a valid LTI system. We note that
realizability holds for arbitrary block values; the restriction to
$GL(n,\mathbb{R})$ serves the group structure, not realization.
Most importantly, this enables the compression of MIMO systems
without a realization gap, preventing the Euclidean drift that
occurs under standard unconstrained methods.

\section{PF-IB OPTIMIZATION VIA LIE ALGEBRA}\label{sec:riemmanianpfib}
The core contribution of this work is the reformulation of the PF-IB objective as a structured optimization problem on the manifold $\mathcal{H}_{GL}$. By utilizing the non-Abelian geometry established in the previous sections, we ensure that the predictive operator $H(\beta)$ remains exactly realizable as a MIMO system throughout the compression process.

\subsection{Optimization Choice: Algebra vs. Manifold}
For practical implementation, we must choose whether to update the
Hankel matrix directly on the manifold $\mathcal{H}_{GL}$ or within
its Lie algebra $\mathfrak{h}_{GL}$. The standard Riemannian
approach works on the manifold: project the Euclidean gradient onto
the tangent space, step, and retract back via the exponential map,
guaranteeing every iterate remains in the group. While
mathematically intuitive, this manifold-centric update faces a
significant hurdle with the PF-IB objective.

Under standard Gaussian assumptions, the PF-IB objective achieves compression by driving the singular values of the predictive operator toward zero. In a manifold-centric framework, this ``squeezing'' of the system's informational mapping forces the operator $H$ toward regions of rank-deficiency. As these singular values diminish, the logarithmic map $\log_{\mathfrak{h}_{GL}}(H)$, which is required to pull gradients from the manifold back to the tangent space, diverges toward negative infinity. This numerical degeneracy causes the optimization to collapse as the tangent space coordinates become unbounded, effectively breaking the geometric link between the manifold and its algebra before an information-theoretic optimum can be reached.

To ensure numerical stability, we adopt a direct algebraic optimization strategy. Instead of updating the matrix $H$ on the manifold, we optimize its coordinates directly within the Lie algebra $\mathfrak{h}_{GL}$  and map the result to the manifold via $H = \exp_{\mathcal{H}_{GL}}(h)$. Because the Lie algebra is a linear vector space, it remains numerically well-defined across all scales of the bottleneck parameter $\beta$. This allows the optimization to capture the necessary predictive information, while the exponential map ensures that every iteration remains full-rank and exactly realizable.
\begin{algorithm}[t]
  \caption{Direct Algebraic Optimization on $\mathfrak{h}_{GL}$}
  \label{alg:pfib-optimization}
  \begin{algorithmic}[1]
    \STATE \textbf{Input:} Initial uncompressed Hankel matrix $H$, iterations $N$, learning rate $\eta$
    \STATE $h(\beta) = \log_{\mathcal{H}_{GL}}(H)$ \COMMENT{Initial map to algebra, Eq.~\eqref{eq:log_map}}
    \FOR{$i = 1$ to $N$}
        \STATE $H(\beta)_i = \exp_{\mathcal{H}_{GL}}(h(\beta)_i)$ \COMMENT{Retract to manifold, Eq.~\eqref{eq:exp_map}}
        \STATE $\nabla_h \mathcal{L} = \nabla_H\mathcal{L}(H(\beta)_i) \odot H(\beta)_i^{\top}$ \COMMENT{Loss gradient in algebra coordinates, Eq.~\eqref{eq:algebra_gradient}}
        \STATE $h(\beta)_{i+1} = h(\beta)_i - \eta\, \mathcal{P}_{\mathfrak{h}_{GL}}\left(\nabla_{h} \mathcal{L}\right)$ \COMMENT{Linear update on $\mathfrak{h}_{GL}$, Eq.~\eqref{eq:linear_update}}
    \ENDFOR
    \STATE \textbf{Output:} $H(\beta)^* = \exp_{\mathcal{H}_{GL}}(h(\beta)^*)$ \COMMENT{Final exactly realizable system}
  \end{algorithmic}
\end{algorithm}


\subsection{Direct Algebraic Optimization}

To optimize the PF-IB objective directly within the stable
linear space $\mathfrak{h}_{GL}$, we derive the gradient of
$\mathcal{L}$ with respect to the algebraic coordinates. Given
the mapping $H(\beta) = \exp_{\mathcal{H}_{GL}}(h(\beta))$
(Eq.~\eqref{eq:exp_map}) with $h(\beta) \in \mathfrak{h}_{GL}$,
the differential of the block-wise matrix exponential admits the
exact Fr\'echet representation
$\delta H_{i,j} = \big(\varphi(\mathrm{ad}_{h_{i,j}})\,
\delta h_{i,j}\big) H_{i,j}$ with
$\varphi(z) = (e^z - 1)/z$
\cite{hall2015lie,higham2008functions}. Retaining the leading
term $\varphi(\mathrm{ad}_{h}) \approx \mathrm{Id}$, i.e.,
$\delta H \approx \delta h \odot H$, the chain rule yields the
block-wise pullback
\begin{equation}\label{eq:algebra_gradient}
    \nabla_{h}\mathcal{L}
    \;\approx\;
    \nabla_{H}\mathcal{L} \odot H(\beta)^{\top},
\end{equation}
where $\odot$ denotes the block Hadamard product of
Eq.~\eqref{eq:blockhadamard} applied with the block-wise
transpose $\big(H(\beta)^\top\big)_{i,j} = H_{i,j}^{\top}$. We
denote $\nabla_{h}\mathcal{L} \equiv \frac{d\mathcal{L}}{dh(\beta)}$
and $\nabla_{H}\mathcal{L} \equiv \frac{d\mathcal{L}}{dH(\beta)}$
(Eq.~\eqref{eq:pfib_gradient}). The truncation error is $\tfrac{1}{2}[h,\,\cdot\,]$ at leading order and vanishes identically in the Abelian case: for the scalar Hankel manifold of \cite{AHMAD2024105150} (SISO systems), Eq.~\eqref{eq:algebra_gradient} holds with equality, and the
neglected terms are precisely the non-commutative corrections that distinguish the block manifold from the scalar case of Sec.~\ref{sec:theoretical_motivation}.

To ensure that the resulting operator maintains the required $\mathfrak{h}_{GL}$ structure, we apply the block-wise anti-diagonal averaging operator $\mathcal{P}_{\mathfrak{h}_{GL}}$ (Eq. \eqref{eq:projection_operator}) to the algebraic gradient. This step projects the raw gradient $\nabla_{{h}} \mathcal{L}$ onto the subspace of $\mathfrak{h}_{GL}$. 

The final update rule for the Lie algebra coordinates at iteration $t$ is given by:
\begin{equation}\label{eq:linear_update}
{h}_{t+1} = {h}_{t} - \eta \mathcal{P}_{\mathfrak{h}_{GL}} \left( \nabla_{{h}} \mathcal{L}\right)
\end{equation}
where $\eta$ is the learning rate. It is important to note that we use the uncompressed Hankel matrix of the original system (Eq. \eqref{eq:hankelmatrix}). This choice guarantees that the
optimization starts at an exactly realizable interior point of $\mathcal{H}_{GL}$, where the logarithmic map (Eq. \eqref{eq:log_map}) is well-defined. The full iterative procedure for this stable Riemannian descent is summarized in Alg. \ref{alg:pfib-optimization}.

By performing the update directly in the algebra $\mathfrak{h}_{GL}$, we leverage a linear coordinate system that is inherently robust to the block degeneracy that arises under strong compression, where $\det(H(\beta)) \xrightarrow{} 0$ and the logarithmic map becomes undefined. The resulting operator $H_{t+1} = \exp_{\mathcal{H}_{GL}}({h}_{t+1})$ is guaranteed to be a strictly full-rank, exactly realizable.

\begin{figure}[t]
\centering
\includegraphics[width=1\linewidth]{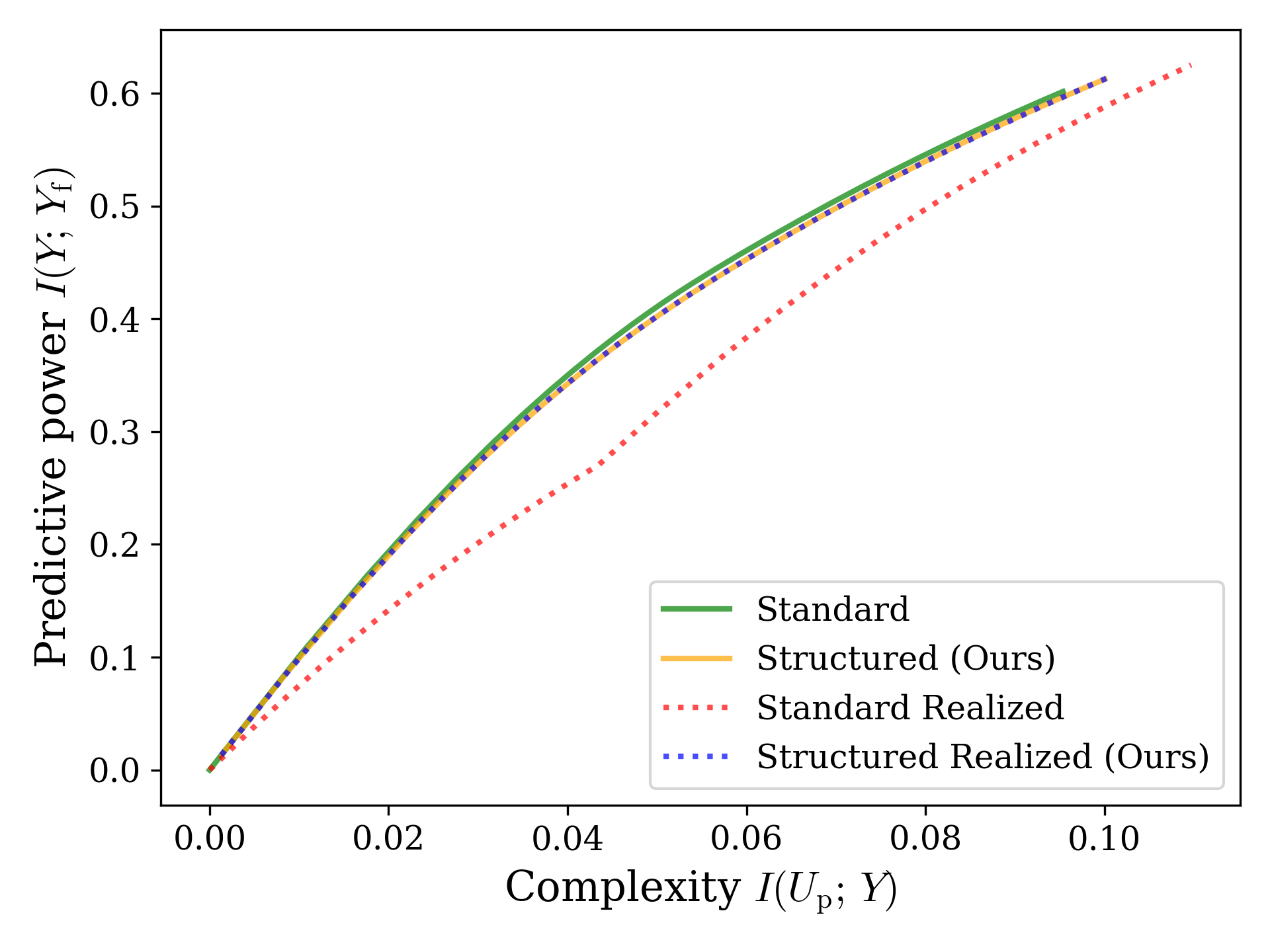}
\caption{The Information Curve showing the tradeoff between complexity $I(\Up;\Yfh)$ and predictive power $I(\Yfh; \Yf)$. The realization gap is visually represented by the vertical distance between the theoretical optimum (standard) and the performance actually achieved by it's approximated realization (standard realized).}
\label{fig:infocurve}
\end{figure}
\begin{table}[t]
    \centering
    \caption{Trajectory Error Comparison}
    \label{tab:trajectory_error_beta}
    \small 
    \begin{tabular*}{\columnwidth}{@{\extracolsep{\fill}} l ccc ccc @{}}
        \toprule
        & \multicolumn{3}{c}{$\beta = 2.06$} & \multicolumn{3}{c}{$\beta = 2.40$}\\
        \cmidrule(lr){2-4} \cmidrule(lr){5-7}
        \textbf{Method} & \textbf{RMS} & \textbf{Max} & \textbf{Std} & \textbf{RMS} & \textbf{Max} & \textbf{Std} \\
        \midrule
        Standard  & 8.33 & 21.54 & 5.41 & 5.65 & 13.52 & 3.37 \\
        \textbf{Structured}  &\textbf{ 1.59} & \textbf{2.77} &\textbf{ 0.62 }& \textbf{1.36} & \textbf{2.44} & \textbf{0.58}\\
        \bottomrule
    \end{tabular*}
\end{table}
\section{EXPERIMENTAL RESULTS}\label{sec:results}
In this section, we evaluate our work with these {\bf R}esearch {\bf Q}uestions:

\begin{enumerate}
    \item \textbf{RQ1 Realizability Gap:} Does the optimization maintain realizability?
    \item \textbf{RQ2 State Estimation:} Do PF-IB representations preserve information relevant for long-horizon prediction?
    \item \textbf{RQ3 Control Performance:} Do PF-IB representations retain information relevant for downstream control tasks?
\end{enumerate}

\begin{figure*}[t]
\centering
\begin{subfigure}[b]{0.24\textwidth}
\centering
\includegraphics[width=\textwidth]{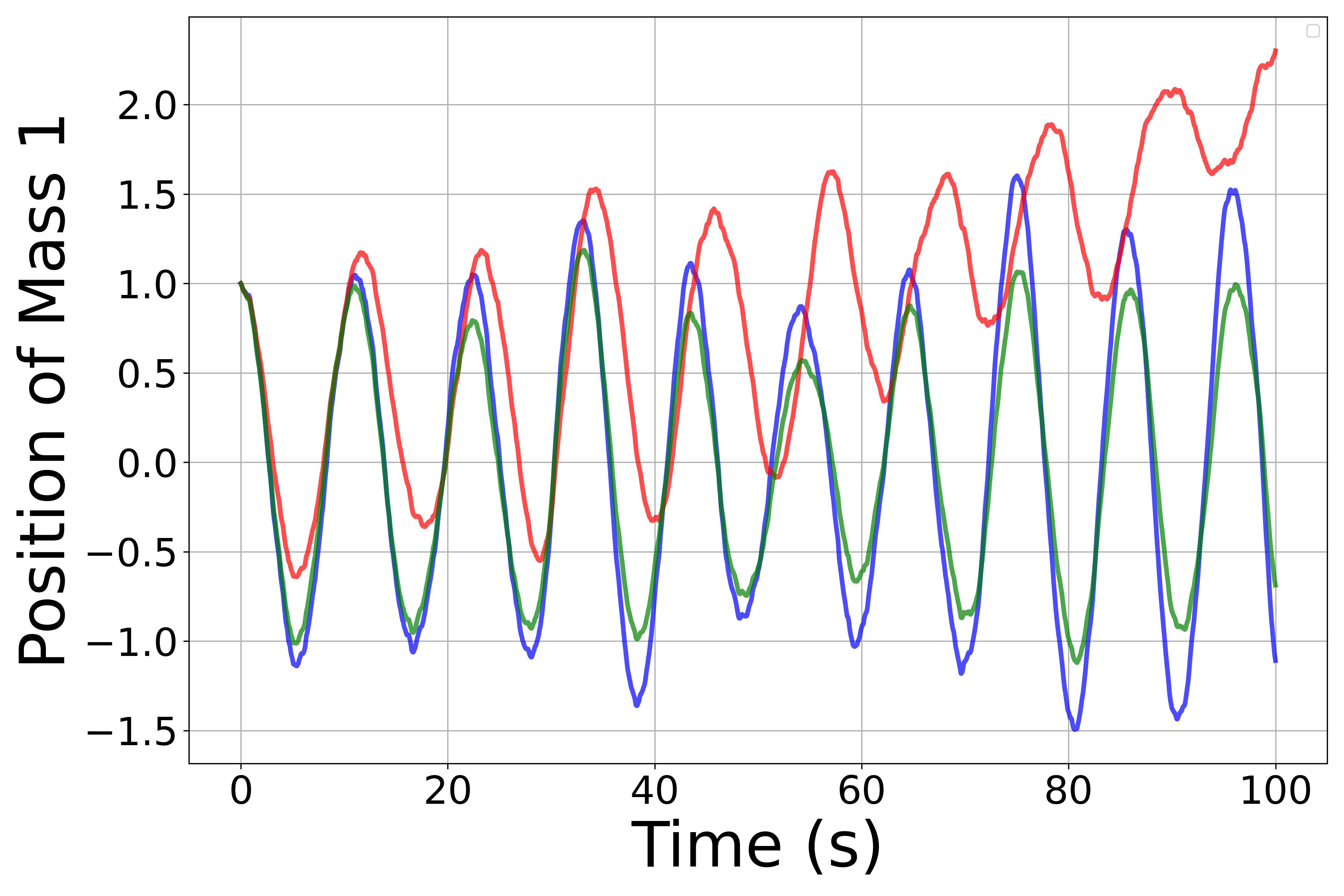}
\caption{}
\end{subfigure}
\hfill
\begin{subfigure}[b]{0.24\textwidth}
\centering
\includegraphics[width=\textwidth]{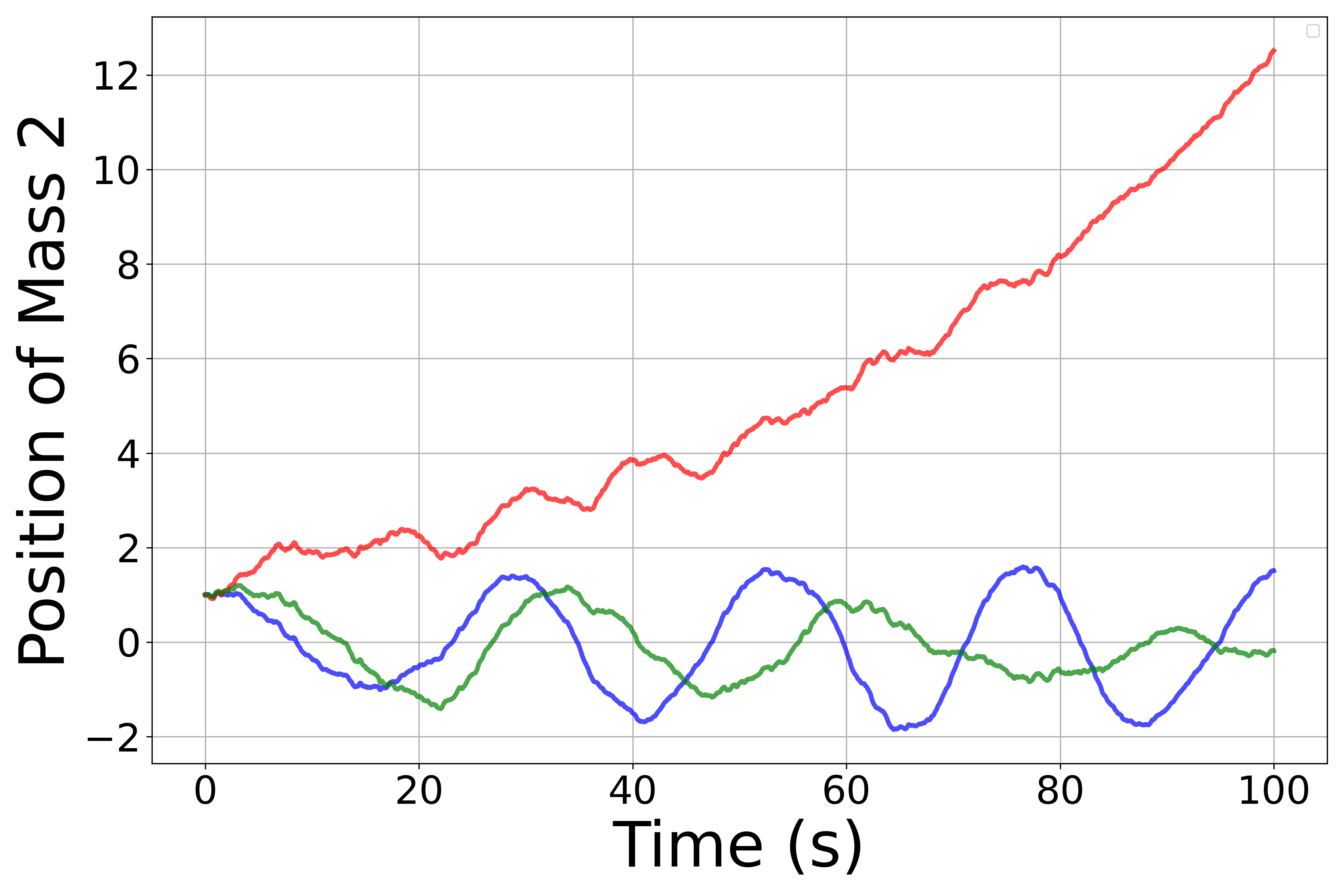}
\caption{}
\end{subfigure}
\hfill
\begin{subfigure}[b]{0.24\textwidth}
\centering
\includegraphics[width=\textwidth]{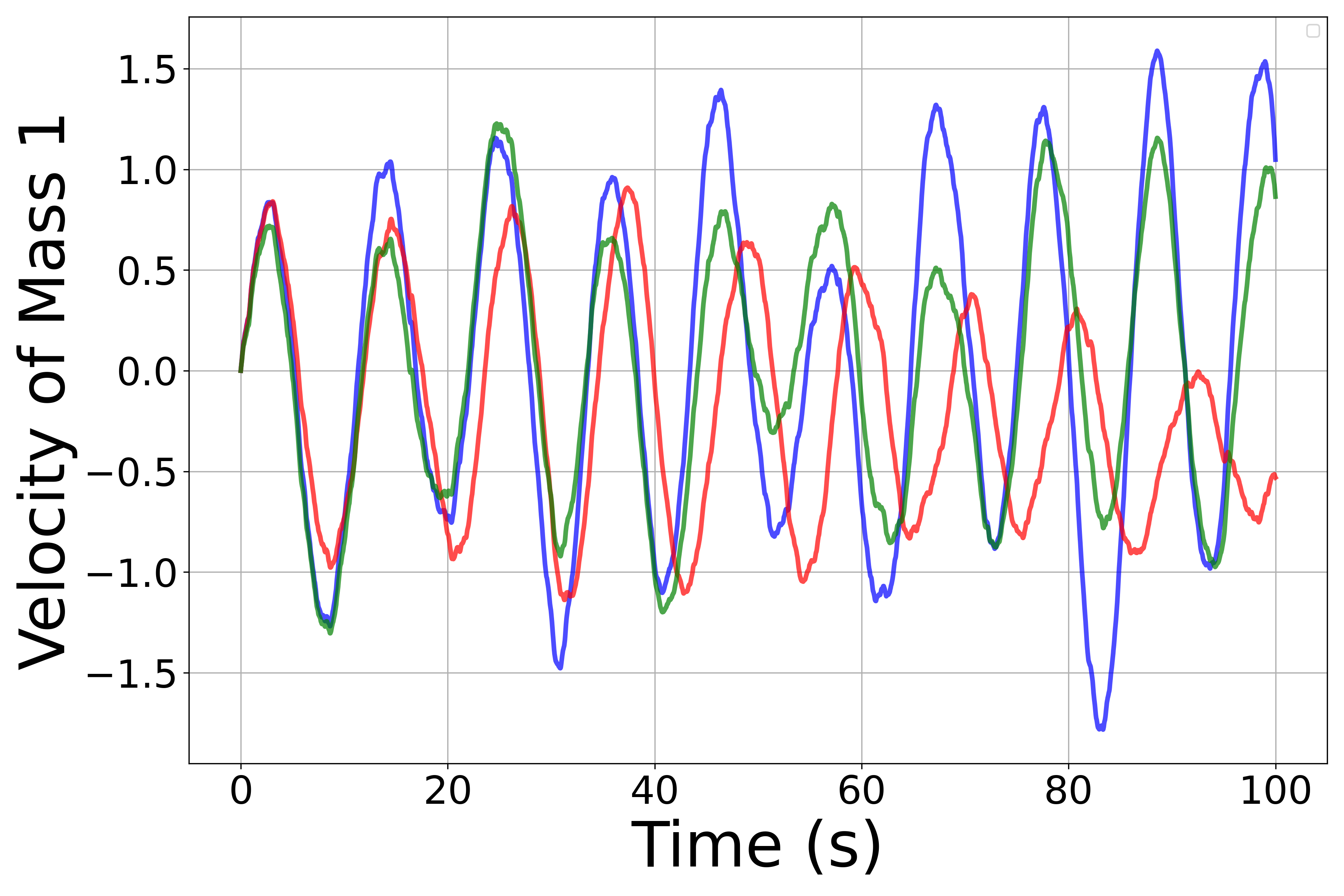}
\caption{}
\end{subfigure}
\hfill
\begin{subfigure}[b]{0.24\textwidth}
\centering
\includegraphics[width=\textwidth]{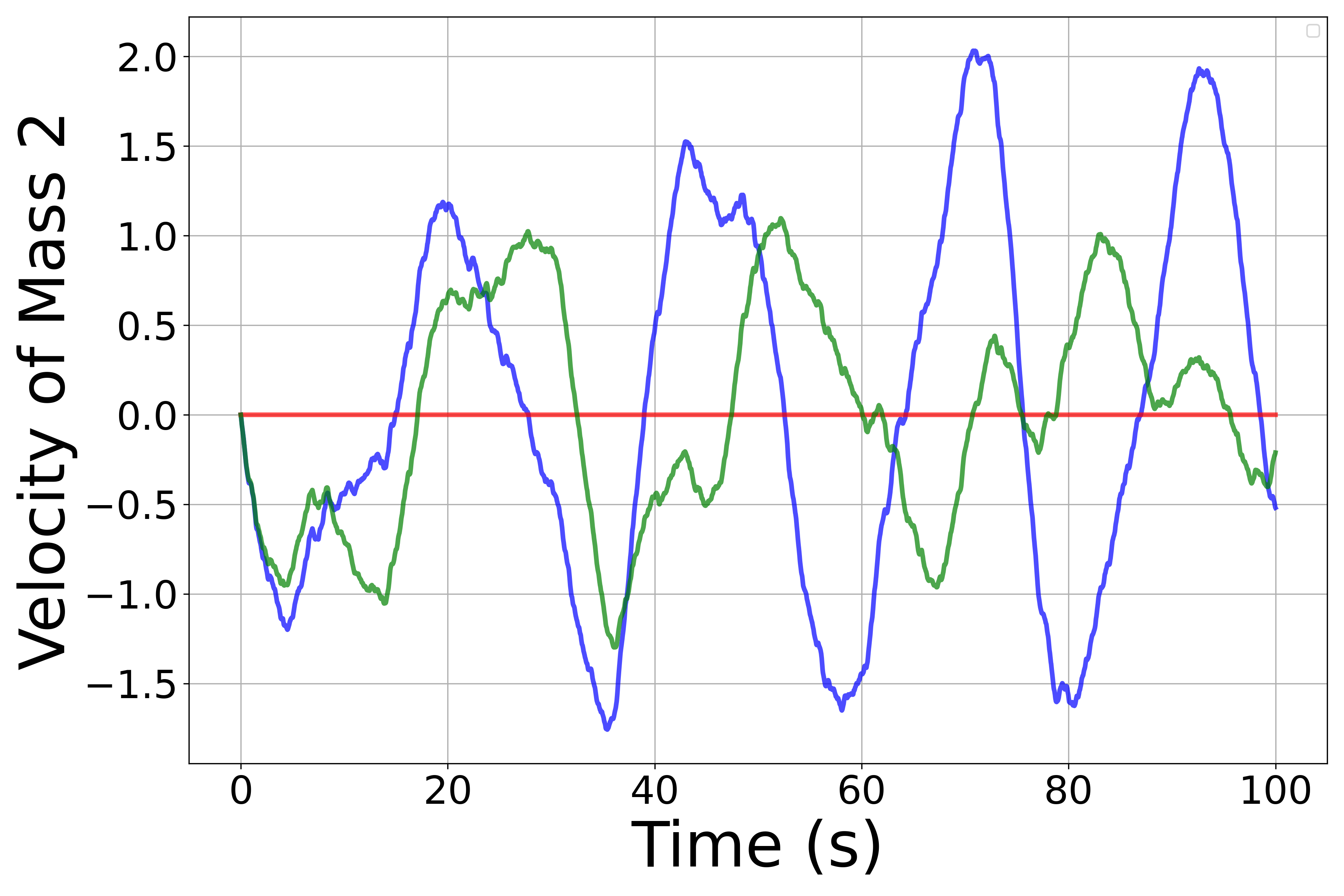}
\caption{}
\end{subfigure}
\caption{Predicted vs. ground truth trajectories for the spring-mass system across different states. {\color{blue}Blue} is ground truth, {\color{OliveGreen}green} is structured realized prediction, and {\color{red}red} is unstructured realized (approximated) prediction.}
\label{fig:trajectories_all}
\end{figure*}
\subsection{Experimental Setting}
We use the same experimental settings as the previous works in  PF-IB \cite{Creutzig2009,7403120} to evaluate our method. Specifically, we consider a two-mass, spring-mass system, which serves as a classic linear dynamical systems. The system consists of two masses ($m_1, m_2$) connected by springs ($k_1, k_2, k_3$) and dampers ($c_1, c_2$), with the state-space representation defined as:
\begin{align}
A =
\begin{bmatrix}
0 & 0 & 1 & 0 \\
0 & 0 & 0 & 1 \\
-\frac{k_1+k_2}{m_1} & \frac{k_2}{m_1} & -\frac{c_1}{m_1} & 0 \\
\frac{k2}{m_1} & -\frac{k_2+k_3}{m_1} & 0 & -\frac{c_2}{m_2}
\end{bmatrix},\\
B = \begin{bmatrix}
0 & 0\\
0 & 0\\
\frac{1}{m_1}& 0\\
0 & \frac{1}{m_2}\\
\end{bmatrix},
C = \begin{bmatrix}
1 & 0 & 0 & 0\\
0 & 1 & 0 & 0\\
\end{bmatrix},
\end{align}

Consistent with the benchmark settings, we set the physical constants to $m_1 = 5, m_2 = \sqrt{15}$, $k_1 = 1, k_2 = 0.5 , k_3= 0$, and $c_1=c_2 = 0$. The system has four states with $p = q = 2$ inputs/outputs, so the
Markov parameters $G_i \in \mathbb{R}^{2\times2}$ form blocks of size
$n = 2$; with past/future horizons of $k = 7$ steps,
$H \in \mathbb{R}^{14 \times 14}$ and the manifold dimension is
$(2k-1)n^2 = 52$.

\subsection{Experimental Result}
\noindent
\textbf{Realizability Gap:} To address RQ1, we evaluate the Realizability Gap by comparing compression performance on the information curve seen in Fig. \ref{fig:infocurve}, which illustrates the trade-off between representation complexity $I(\Up;\Yfh)$ and predictive power $I(\Yfh; \Yf)$. We observe that while unstructured Euclidean optimization (Standard) achieves a high theoretical bound, there is a significant information drop when approximated into a realizable structure (Standard realized). In contrast, our Lie-algebraic method (Structured) exhibits no realization gap, as the optimized and realized curves perfectly overlap. Notably, our structured curve remains extremely close to the unstructured upper bound, proving that we achieve near-optimal compression without sacrificing realizability.

\noindent
\textbf{State Estimation}
To evaluate RQ2, we analyze the model's ability to reconstruct long-horizon trajectories from compressed representations. To ensure these trajectories represent realistic physical movement,
we generate five independently randomized trajectories, each driven
by a different realization of a first-order autoregressive (AR(1))
input sequence with correlation coefficient $\rho = 0.9$, producing
temporally correlated actions that prevent high-frequency jitter and
explore the state space in a manner consistent with continuous
physical systems. Both methods are evaluated on the same five randomly generated
input sequences. As shown in Table \ref{tab:trajectory_error_beta}, our method achieves a over 5 times reduction in RMS compared to the baseline at $\beta = 2.06$, a gap that persists even as $\beta$ increases. This improvement is visually evident in Fig. \ref{fig:trajectories_all}. It is important to note that while our structured realization remains stable across all values of $\beta$, we restrict the quantitative comparison plots to $\beta > 2.0$ because the standard baseline realization becomes numerically unstable during reconstruction below this threshold.

\begin{figure}[t]
\centering
\begin{subfigure}[b]{0.9\linewidth}
\centering
\includegraphics[width=\textwidth]{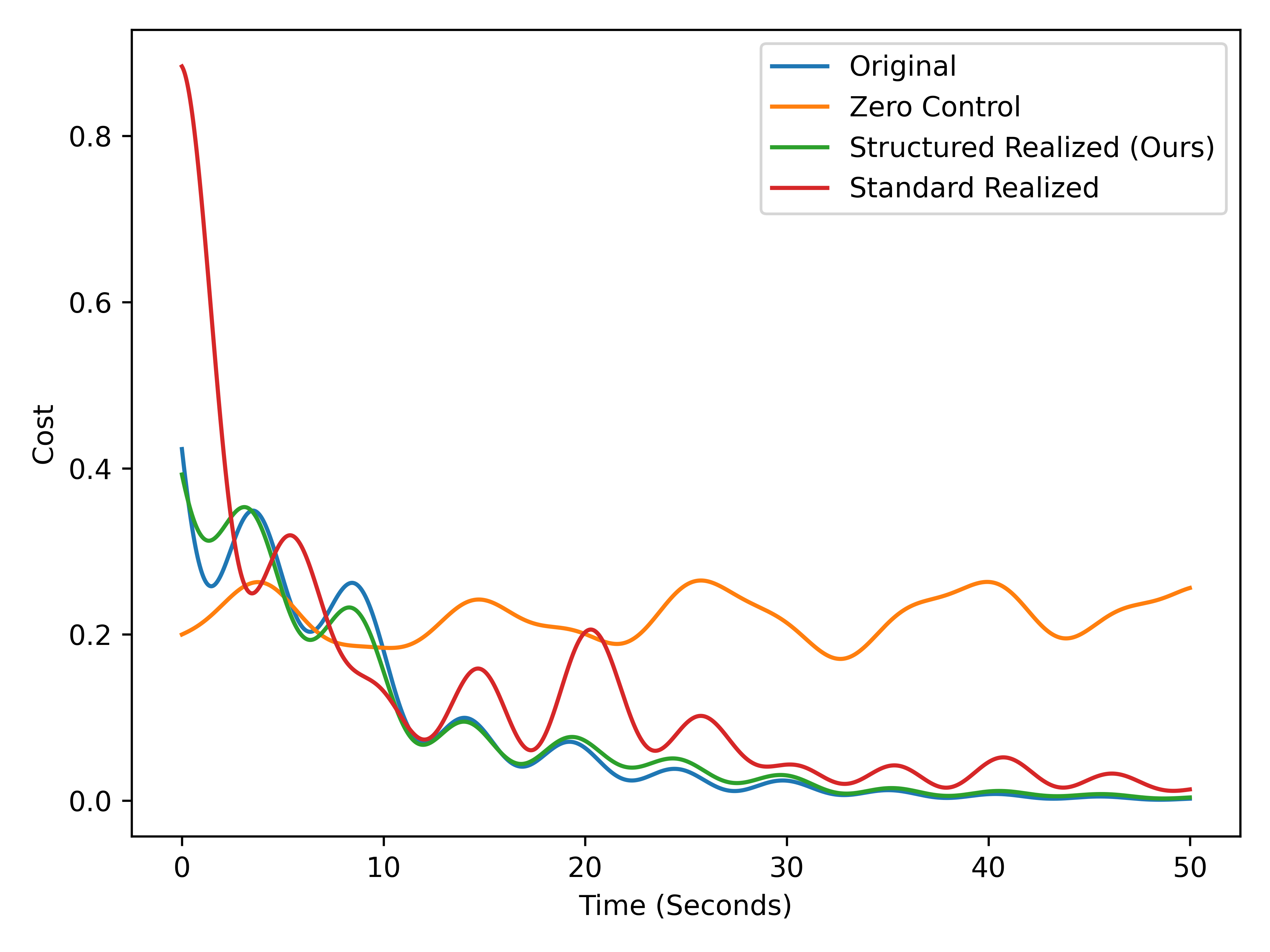}
\caption{LQR Cost Comparison}
\end{subfigure}
\hfill
\begin{subfigure}[b]{0.9\linewidth}
\centering
\includegraphics[width=\textwidth]{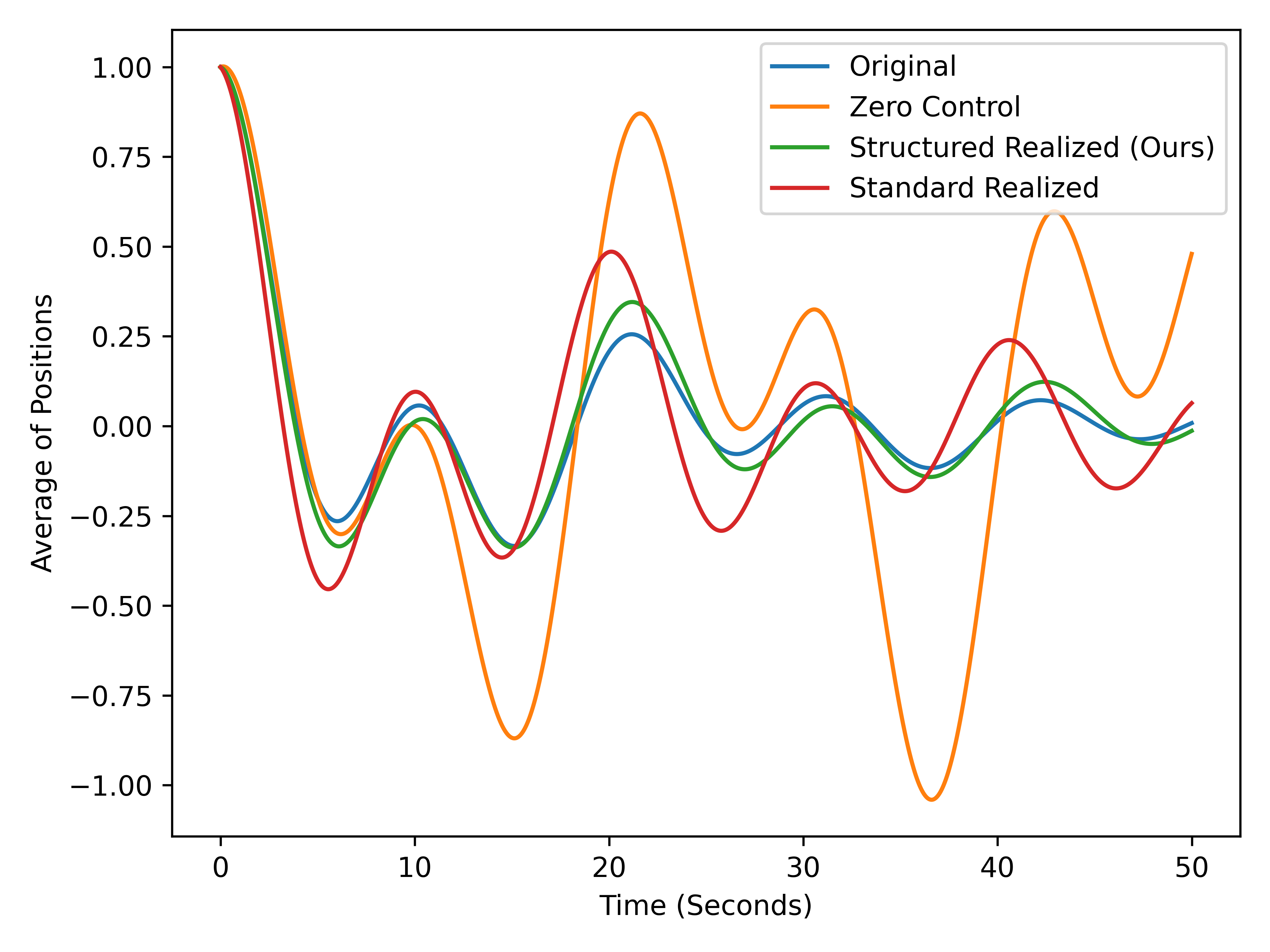}
\caption{LQR Trajectories}
\end{subfigure}
\caption{Downstream Control Performance using PF-IB representations.}
\label{fig:lqr_results}
\end{figure}
\noindent

\textbf{Control Performance:}
Finally, we address RQ3 by evaluating the compressed
representations in a downstream Linear Quadratic Regulator (LQR)
task. For each method, we first extract a realization
$(\hat{A}, \hat{B}, \hat{C})$ from $H(\beta)$ via the Ho--Kalman
construction; this extraction is exact for our structured operator
and approximate for the baseline. We then design a discrete-time
LQR controller on $(\hat{A}, \hat{B})$ with weights $Q = I$ and
$R = 10I$, and apply the resulting gain to the original full-order
plant. The reported cost is therefore the true closed-loop cost on
the original system, and since both methods use identical weights,
differences in Fig.~\ref{fig:lqr_results} reflect realization
fidelity alone. In the figure, \emph{Original} denotes the LQR
controller synthesized directly on the full-order system, serving
as the achievable optimum, and \emph{Zero Control} denotes the
uncontrolled open-loop response.

Our structured realization maintains performance near this optimum,
whereas the standard realization diverges significantly. As in the
state estimation task, we restrict the comparison to $\beta > 2.0$,
below which the standard realization becomes numerically unstable
while ours remains stable throughout.

\section{CONCLUSIONS AND FUTURE WORKS}\label{sec:conclusion}
We addressed the realization gap in information-theoretic system simplification by constraining PF-IB to the block Hankel manifold. By formalizing these dynamical constraints within a Lie-algebraic framework, we delivered a method that ensures compressed representations remain physically realizable as valid linear systems. Our results demonstrate that by bridging this realization gap, we achieve significantly higher fidelity in state estimation and more stable performance in downstream LQR control tasks compared to standard method.

While this work establishes a rigorous foundation for structured compression, it contains specific limitations that provide avenues for future research. Currently, our theoretical proof that the block Hankel manifold forms a Lie group requires the constituent elements to be invertible (members of the General Linear Group). Relaxing this constraint to include singular matrices would further generalize the framework. Additionally, while this study focused on linear systems, extending these Lie-algebraic constraints to non-linear dynamics represents a critical next step toward scaling these benefits to more complex, real-world systems.

Beyond the specific PF-IB setting, our results suggest a broader
principle: information-theoretic compression and structural
realizability need not trade off against each other when the
structure is encoded as the geometry of the optimization domain
rather than enforced by post-hoc projection. The block Hankel
manifold is one instance; the same recipe, identifying the
algebraic structure of a model class, establishing its Lie group
geometry, and optimizing in the associated algebra, applies to
other structured model classes in identification and data-driven
control. Because the compressed operator is realizable at every
iterate, the approach also composes naturally with downstream
certificates: any estimator or controller synthesized from an
iterate inherits a valid state-space model, at any stopping point.

\section{APPENDIX}\label{sec:Appendix}
We prove that $\mathcal{H}_{GL}$ forms a lie group under the block hadamard product. To formally establish $\mathcal{H}_{GL}$ as a Lie group, we must satisfy the following criteria:

\noindent\textbf{Group Axioms (Algebraic):}
\begin{itemize}
\item \textbf{Closure:} For any $A, B \in \mathcal{H}_{GL}$, the product $A \odot B$ must also be in $\mathcal{H}_{GL}$.
\item \textbf{Associativity:} For all $A, B, C \in \mathcal{H}_{GL}$, $(A \odot B) \odot C = A \odot (B \odot C)$.
\item \textbf{Identity:} There must exist an element $E \in \mathcal{H}_{GL}$ such that $A \odot E = E \odot A = A$ for all $A$.
\item \textbf{Inverses:} For every $A \in \mathcal{H}_{GL}$, there must exist an element $A_{inv} \in \mathcal{H}_{GL}$ such that $A \odot A_{inv} = E$.
\end{itemize}
\noindent\textbf{Manifold Structure (Geometric):}
\begin{itemize}
\item \textbf{Smoothness:} The set $\mathcal{H}_{GL}$ must locally resemble Euclidean space $\mathbb{R}^D$ and possess smooth group operations (multiplication and inversion) that are infinitely differentiable ($C^\infty$).
\item \textbf{Dimension:} The manifold must possess a well-defined, constant dimension corresponding to the total degrees of freedom inherent in the block Hankel structure.
\end{itemize}

\begin{theorem}
The set $\mathcal{H}_{GL}$ under the block Hadamard product $\odot$ forms a Lie group over $\mathbb{R}$.
\end{theorem}

\textbf{Proof 3.1 Closure} Let $A, B \in \mathcal{H}_{GL}$ be two block Hankel matrices of size $k \times k$ blocks, where each block $A_{i,j}, B_{i,j} \in GL(n, \mathbb{R})$. We define the product $C = A \odot B$ via the block Hadamard product as specified in Eq. \eqref{eq:blockhadamard}. To prove closure, we must show that $C \in \mathcal{H}_{GL}$.

First, we verify the invertibility of the resulting blocks. Since $GL(n, \mathbb{R})$ is a group under matrix multiplication, the product of any two invertible matrices $A_{i,j}$ and $B_{i,j}$ is itself an invertible matrix. Thus, $C_{i,j} = A_{i,j} B_{i,j} \in GL(n, \mathbb{R})$ for all $i,j$.

Second, we verify the Hankel structure. A matrix is block Hankel if its block entries are constant along its anti-diagonals, i.e., $A_{i,j} = A_{i+1,j-1}$ and $B_{i,j} = B_{i+1,j-1}$ for all valid $i,j$. Consider the $(i+1,j-1)$-th block of $C$:
\begin{align}
C_{i+1,j-1} &= A_{i+1,j-1} B_{i+1,j-1} = A_{i,j} B_{i,j} = C_{i,j}
\end{align}
This algebraic identity holds because the individual blocks $A_{i,j}$ and $B_{i,j}$ satisfy the Hankel property. Since $C$ has constant values along its anti-diagonals and all blocks remain in $GL(n, \mathbb{R})$, it follows that $C \in \mathcal{H}_{GL}$. Therefore, the set $\mathcal{H}_{GL}$ is closed under the block Hadamard product.

Finally, we verify that $C$ retains the uniform block structure. Since
$\mathcal{H}_{GL}$ is defined with respect to a fixed $k \times k$
partition into $n \times n$ blocks (Sec.~IV-A), and the block Hadamard
product of Eq.~(6) acts on this same grid, each block
$C_{i,j} = A_{i,j}B_{i,j}$ is the product of two $n \times n$ matrices
and is therefore itself $n \times n$. Hence all blocks of $C$ along
every anti-diagonal have identical size, and $C$ is a block Hankel
matrix on the same partition as $A$ and $B$, so $C \in \mathcal{H}_{GL}$. 

\textbf{Proof 3.2 Associativity} Let $A, B, C \in \mathcal{H}_{GL}$ be three block Hankel matrices of size $k \times k$ blocks, where each block is of size $n \times n$. We must show that $(A \odot B) \odot C = A \odot (B \odot C)$.

Recall that the block Hadamard product is defined block-wise. By applying the definition to the $(i,j)$-th block entry, we have:
\begin{align}
\left[ (A \odot B) \odot C \right]_{i,j} &= (A \odot B)_{i,j} C_{i,j} = (A_{i,j} B_{i,j}) C_{i,j}
\end{align}
Since the individual blocks $A_{i,j}, B_{i,j}, C_{i,j}$ are elements of $GL(n, \mathbb{R})$, they satisfy the standard associative property of matrix multiplication: $(A_{i,j} B_{i,j}) C_{i,j} = A_{i,j} (B_{i,j} C_{i,j})$. Therefore:
\begin{align}
\left[ (A \odot B) \odot C \right]_{i,j} &= A_{i,j} (B_{i,j} C_{i,j}) \\
= A_{i,j} (B \odot C)_{i,j} &= \left[ A \odot (B \odot C) \right]_{i,j}
\end{align}
Since this equality holds for all block indices $i, j \in \{1, \dots, k\}$, it follows that $(A \odot B) \odot C = A \odot (B \odot C)$. Thus, the block Hadamard product is associative for all elements in $\mathcal{H}_{GL}$.

\textbf{Proof 3.3 Identity} To show the existence of an identity element, we must find a matrix $E \in \mathcal{H}_{GL}$ such that for all $A \in \mathcal{H}_{GL}$, $A \odot E = E \odot A = A$.

Let $A \in \mathcal{H}_{GL}$ be a block Hankel matrix of $k \times k$ blocks of size $n \times n$. We define $E$ as the $kn \times kn$ matrix where every block entry is the $n \times n$ identity matrix $I$:

\begin{equation}
E =\begin{bmatrix}I & I & \cdots & I \\
I & I & \cdots & I \\
\vdots & \vdots &\ddots & \vdots \\ 
I & I & \cdots & I\end{bmatrix}.
\end{equation}

Since every block in $E$ is the identity matrix $I$, and $I$ is constant along all anti-diagonals, $E$ satisfies the block Hankel property. Furthermore, since $\det(I) = 1$, each block is an element of $GL(n, \mathbb{R})$. Thus, $E \in \mathcal{H}_{GL}$.

Next, we verify the identity property by considering the $(i,j)$-th block entry of the product $A \odot E$:
\begin{align}
(A \odot E)_{i,j} &= A_{i,j} E_{i,j} = A_{i,j} I = A_{i,j}.
\end{align}
By the same logic, $(E \odot A)_{i,j} = I A_{i,j} = A_{i,j}$. Since this holds for all $i,j \in \{1, \dots, k\}$, we have $A \odot E = E \odot A = A$. This demonstrates that $E$ is the unique identity element for the group $(\mathcal{H}_{GL}, \odot)$.

\textbf{Proof 3.4 Inverses} To show the existence of an inverse for each element, we must demonstrate that for any block Hankel matrix $A \in \mathcal{H}_{GL}$, there exists an inverse $A_{inv} \in \mathcal{H}_{GL}$ such that $A \odot A_{inv} = E$.

Let $A \in \mathcal{H}_{GL}$ be a block Hankel matrix of $k \times k$ blocks of size $n \times n$. We define the inverse matrix $A_{inv}$ by taking the matrix inverse of each block entry:

\begin{equation}
(A_{inv})_{i,j} = (A_{i,j})^{-1}.
\end{equation}

First, we observe that since $A \in \mathcal{H}_{GL}$, every block $A_{i,j}$ is an element of $GL(n, \mathbb{R})$. It follows from the group properties of $GL(n, \mathbb{R})$ that the matrix inverse $(A_{i,j})^{-1}$ exists and is also an element of $GL(n, \mathbb{R})$.

Next, we verify the Hankel structure of the inverse. Since $A$ is a block Hankel matrix, we have $A_{i,j} = A_{i+1,j-1}$ for all valid $i, j$. Therefore:
\begin{align}
(A_{inv})_{i+1,j-1} &= (A_{i+1,j-1})^{-1} = (A_{i,j})^{-1} = (A_{inv})_{i,j}.
\end{align}
This shows that $A_{inv}$ has constant values along its anti-diagonals and is therefore a block Hankel matrix. Since each block remains in $GL(n, \mathbb{R})$, we confirm $A_{inv} \in \mathcal{H}_{GL}$.

Finally, we verify that $A \odot A_{inv} = E$ by considering the $(i,j)$-th block entry:
\begin{align}
(A \odot A_{inv})_{i,j} &= A_{i,j} (A_{inv})_{i,j} = A_{i,j} (A_{i,j})^{-1} = I.
\end{align}
Because this results in the identity matrix $I$ for all blocks, we have $A \odot A^{-1} = E$. Thus, every element in $\mathcal{H}_{GL}$ possesses a unique inverse within the set.

\textbf{Proof 3.5 Smoothness and Topology} To prove that $\mathcal{H}_{GL}$ is a Lie group, we must show that the group operations, multiplication ($\odot$), and inversion ($(\cdot)_{inv}$), are smooth ($C^\infty$) maps.

First, we identify the topological structure. The General Linear Group $GL(n, \mathbb{R})$ is an open subset of the vector space $\text{Mat}(n, \mathbb{R})$, defined by the non-vanishing of the continuous determinant function $\det(X) \neq 0$. Since $\mathcal{H}_{GL}$ is a finite collection of such blocks restricted by linear Hankel constraints, it inherits this property, forming an open submanifold of the space of block Hankel matrices. This ensures that every point in $\mathcal{H}_{GL}$ has a neighborhood diffeomorphic to Euclidean space $\mathbb{R}^D$, satisfying the manifold requirement.

Next, we verify the smoothness of the group operations. The block Hadamard product $A \odot B$ is defined by the block-wise matrix multiplication of its entries. Since matrix multiplication is a polynomial operation in the entries of the matrices, the map $\odot: \mathcal{H}_{GL} \times \mathcal{H}_{GL} \to \mathcal{H}_{GL}$ is $C^\infty$. Similarly, inversion map $A \mapsto A_{inv}$ is defined by the block-wise matrix inversion $(A_{i,j})^{-1}$. Matrix inversion is a smooth map on the open set $GL(n, \mathbb{R})$, as its entries are rational functions (via Cramer's rule) with non-zero denominators ($\det(A_{i,j}) \neq 0$); thus, the inversion map is also $C^\infty$.

Since the set is a smooth manifold and the group operations are smooth, $\mathcal{H}_{GL}$ is a Lie group.

\textbf{Proof 3.6 Dimension} We must determine the constant dimension $D$ of the manifold $\mathcal{H}_{GL}$. The dimension is defined by the number of independent real parameters required to uniquely specify an element in the set.

First, we recall that a block Hankel matrix with $k \times k$ blocks is uniquely determined by its first row and last column. This results in $2k-1$ unique block entries.

Next, we account for the internal structure of each block. Each unique block $G_m$ is an element of $GL(n, \mathbb{R})$. Since $GL(n, \mathbb{R})$ is an open subset of $\mathbb{R}^{n \times n}$, it inherits the dimension of the space of $n \times n$ matrices, which is $n^2$. Therefore, the total number of independent parameters (the dimension of the manifold) is:

\begin{equation}
\text{dim}(\mathcal{H}_{GL}) = (2k - 1)n^2.
\end{equation}
Since $k$ and $n$ are fixed parameters defining the space $\mathcal{H}_{GL}$, this dimension is constant across the entire manifold.




\bibliographystyle{IEEEtran}
\bibliography{references}
\end{document}